# *Three-Dimensional Chiral MetaCrystals*

*Marco Esposito ∥[1*], Mariachiara Manoccio ∥[1], Angelo Leo[1], Massimo Cuscunà[1], Yali Sun[2], Eduard Ageev[2], Dmitry Zuev[2*], Alessio Benedetti[3], Iolena Tarantini[4], Adriana Passaseo[1] and Vittorianna Tasco[1*]*

1. CNR NANOTEC Institute of Nanotechnology, Via Monteroni, 73100 Lecce, Italy
2. ITMO University, Department of Physics and Engineering, 49 Kronverkskiy av., St. Petersburg 197101, Russia
3. D.I.E.T. Department, "Sapienza: Università di Roma", Via Eudossiana 18, I-00184, Rome, Italy.
4. Department of Mathematics and Physics Ennio De Giorgi, University of Salento, Via Arnesano, 73100 Lecce.

\* Corresponding authors e-mail: *marco.esposito@nanotec.cnr.it; vittorianna.tasco@nanotec.cnr.it; d.zuev@metalab.ifmo.ru*

∥ these authors contributed equally to the work

**Abstract**
Fine control of the chiral light-matter interaction at the nanoscale, by exploiting designed metamaterial architecture, represents a cutting-edge craft in the field of biosensing, quantum and classic nanophotonics. Recently, artificially engineered 3D nanohelices have demonstrated programmable wide chiroptical properties by tuning materials and architecture, but fundamental diffractive aspects that are to the origin of chiral resonances still remain elusive. Here, we proposed a novel concept of three-dimensional chiral MetaCrystal, where the chiroptical properties are finely tuned by in-plane and out-of-plane diffractive coupling. Different chiral dipolar modes can be excited along the helix arms, generating far field optical resonances and radiation pattern with in-plane side lobes and suggesting that a combination of efficient dipole excitation and diffractive coupling matching controls the collective oscillations among the neighbor helices in the chiral MetaCrystal. This concept enables the tailorability of chiral properties in a broad spectral range for a plethora of forefront applications, since the proposed compact chiral MetaCrystal can be suitable for integration with quantum emitters and can open perspectives in novel schemes of enantiomeric detection.

The continuously growing interest towards metamaterials development comes from the possibility to enable non-naturally occurring optical properties[1,2]. In the last years metamaterials have been realized exploiting different materials, shapes, layouts and dimensions to exhibit functional and exotic optical properties in various operational regions, from infrared to visible, for several applications of technological interest[3–5].

Fundamentally, the optical response of a metamaterial is tailored by the interplay between constitutive material and geometric architecture. When the basic building block possesses parity inversion due to a swirling shape, chiral metamaterials[6–9] could be built, exhibiting anisotropic behavior when interacting with circularly polarized light (CPL). In this frame, chiroptical properties such as circular dichroism (CD) and optical rotation can reach higher orders of magnitude with respect to what observed in naturally occurring chiral molecules[10,11], and, consequently, are of potential interest in many application fields, such as biosensing, and chemistry[12–15]. In particular, chiral metamaterials can represent compact photonic elements as ideal alternative to bulky systems for light polarization handling and to complement photonics integrated circuits[16–19].

The most intuitive and intrinsically chiral shape is the helix geometry. Previous studies on metallic nano and micro helices have shown the dependence of far field optical resonances, as well as of the related radiation pattern, on the interplay between structural sizes and inspecting light wavelength and polarization[20–22] and the possibility to get large chiroptical effects from this family of metamaterial [23–25]. We also know that a fundamental chiral meta-atom can be identified as the smallest helix segment exhibiting non-zero chiroptical effects within a certain spectral range[26]. A further step in such a technology can be the three-dimensional (3D) ordering of helix-shape nanostructures with designed structure and periodicities, to simultaneously control the degree of far field coupling with incident light in the forward direction, and the interference among scattered fields from neighboring helices, as a function of circular polarization. This can lead to Chiral MetaCrystals (CMC), that is 3D arrangements of chiral meta-atoms, acting as a novel class of optical components, with circular polarization-dependent diffractive coupling, driven by both, the out-of-plane and the in-plane parameters of the crystal. Such a spectral and polarization engineering, not achievable at the single nanoparticle level[27], is suitable for easy implementation in compact photonic systems and components, preferably working under normal incidence excitation. On the other hand, a full and effective control on both the CMC rulers (in-plane and out-of-plane lattice constants) poses significant technological challenges, particularly at VIS frequencies. A limited number of nanofabrication techniques offers complete flexibility towards such a multidimensional level of engineering. Even though in the last years helix nanostructures have been demonstrated, with high precision and accuracy, by DNA-assisted synthesis[28,29], by glancing angle deposition[23,30,31], and by focused ion/electron beam processing[32,33], the deterministic construction of a helix-based CMC requires simultaneous control over 3D parameters and over the high precision positioning of seeding points, hardly achievable with most of the available techniques[30].

In the present work, we experimentally demonstrate the concept of CMC composed by metallic nanohelices. We first unveil the physical origin of the optical resonances at the single particle level, leveraged by the out-of-plane lattice constant, by combining numerical studies, based on finite difference time domain tools and realistic material parameters, with far field characterization. Then, we focused on the evolution from single element to collective behavior, taking into account the radiation pattern to estimate and understand mutual helix interaction. Such a thorough investigation is needed to finely engineer the diffractive coupling regime, along the in-plane and out-of-plane directions, and to achieve unprecedented control over the circular polarization discrimination exhibited by the CMC through its 3D lattice parameters, flexibly controlled by the employed focused ion beam processing.

The CMC concept underlying this work is schematically shown in **Figure 1**. The building blocks are metallic nanowires of diameter 120 nm, wound along the vertical direction to create single loop nanohelices, thus consisting of two chiral meta-atoms [26]. In our study, the helix vertical pitch (VP) and length have been varied to get resonant behavior throughout the visible range as an effect of circular polarized light (CPL) excitation (Figure 1-a)[9,26,34]. Moreover, to achieve efficient CP discrimination of transmitted light in the VIS under normal incidence, it is desirable that each helix mainly works in axial mode, but also that collective behavior can be exploited and engineered[35]. Therefore, following RF antenna theory[36,37] and its extension to diffractively coupled helices[22], we engineered single helix geometry in order to have maximum efficiency radiating along the vertical axis, but with secondary components also radiating in the helix plane. This translates into a "quasi axial" radiation pattern, as schematically shown in Figure 1-b: with such a design criterion, secondary side lobes of single helix radiation pattern appear, and their spatial extension, in the hundred nm range, can control the mutual far-field interaction between helices. Therefore, if the designed chiral elements are arranged in a squared array (Figure 1-c), to build up a diamond-structured crystal[38,39], the in-plane lattice parameter (LP), represented by the center-to-center distance among helices, can be chosen to enable diffractive coupling and far field interaction also between the single resonant helices.

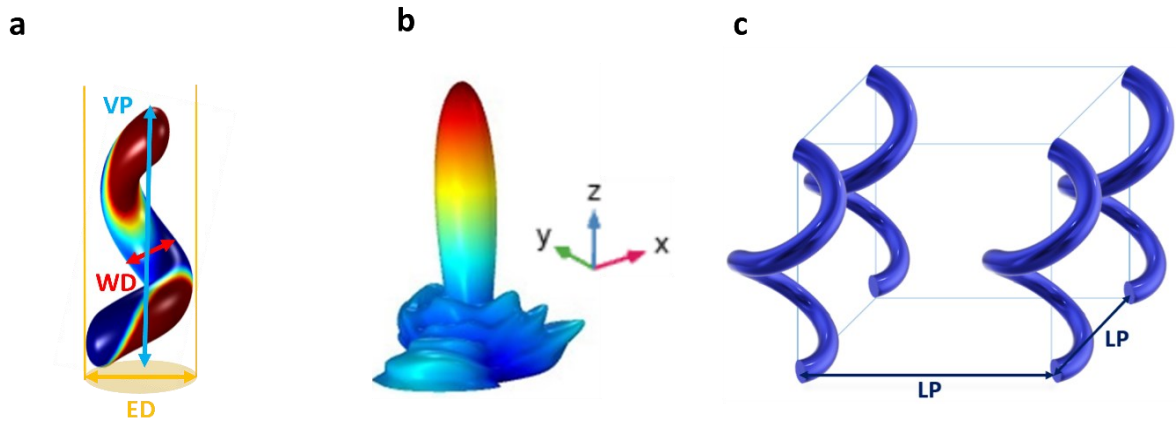

**Figure 1.** Schematic of the CMC concept. a) The unit cell consists of a metallic helix identified by the wire diameter (WD), the external diameter (ED) and the vertical pitch (VP). The spectral response of the single metallic element depends on the number of dipoles excited by inspecting light at a specific wavelength[20]. b) The relationship between ED and inspecting wavelength determines the axial or radial operation of the antenna element. With a chosen ED of 300 nm, in the VIS range, our helices work in a "quasi-axial" scattering regime[36], as schematically shown by the qualitative radiation pattern. Such a radiation profile allows for efficient transmission from the single element along the forward direction, with an additional in plane component enabling collective interaction from the array. c) Center-to-center distance among the helices (lattice parameter LP) represents the in-plane lattice parameter of the complete CMC.

The first step of our analysis was a thorough investigation of the single helix far field response. We focused on the scattering behavior, as a function of circular polarization for different out-of-plane lattice parameters (VP), as shown in Figure 2-a, fabricated by Focused Ion Beam Induced Deposition (FIBID) on ITO/glass substrate (details in Experimental Section). Couples of single enantiomers (right-handed RH and left-handed LH, respectively) with identical geometry and composition were realized. Here, a RH structure is considered, with VP varied to keep a ratio with the excitation wavelength close to 1 (VP= 350 nm, 550 nm and 800 nm)[22]; the total length of the nanowire, $l = \sqrt{(\pi * ED)^2 + VP^2}$, doesn't exceed 1.2 micron in the considered structures. Therefore, according to previous theories [20,40], it is reasonable to assess that the far field response of the single elements can exhibit eigenmodes in the visible spectral range, taking into account the plasmonic losses and the material behavior associated with the realistic dispersion of the platinum-carbon mixture constitutive medium, as commonly obtained with FIBID technology[41].

The numerical calculations of Figure 2-b were performed based on finite element method (FEM) approach (see Experimental Section for further details), and the scattering spectra were normalized to their maxima, considering the experimental set-up described below. First of all, they show that, when incident light is left circularly polarized (LCP), that is, opposite to structure handedness (RH in our case), the scattering resonances are not strongly influenced by the structure length (i.e., by VP, Figure 2-b, left panel). The response appears as a broad peak centered at about 500 nm, without substantial changes in position and modal linewidth with VP in the investigated range. Moreover, as shown by the calculated electric field distribution along the helix surface, these transitions can be related to the excitation of electric dipoles, increasing in number with VP (from n = 3 to n = 5).

On the other hand, when incident CPL is switched to RCP (Figure 2-b, right panel), thus matching the structure handedness (RH), the effect of VP becomes more significant, with the appearance of additional resonances in the whole spectrum, which were not present in the previous cases.

Coupling and hybridization among the different dipoles can arise, governed by the distance between dipoles, that, in turn, can be set acting on VP[42]. For small VP (350 nm), the plasmonic dipoles belong to continuous arms which are close to each other, and this can lead to vertical dipole coupling and to spectrally broad resonance. It is also intuitive that, in this case, the spectral responses under RCP and LCP excitation are very similar since the geometrical chirality is weak, and the helix approaches an achiral hollow cylinder. As VP increases to 550 nm, the two CPLs can clearly excite different modes associated to a different dipole number. In particular, the RCP scattering spectrum exhibits additional and redshifted peaks, where the peak with the lowest number of dipoles corresponds to the mode at longer wavelengths, while the highest order continues to match the resonance generated by opposite handedness exciting light (LCP light). Moreover, the mutual interactions between adjacent helix arms are reduced here, given the increased distance among them. All these features lead to the onset of well-defined resonances in the RCP case and to multiple bands of circular polarization discrimination.

When the height is increased up to 800 nm, the low-n plasmonic resonances are further redshifted, clearly unveiling the excited high-n optical resonances. Interaction between arms is further reduced here, as well as the related plasmonic noise due to the stacked dipole weak coupling. The helix is stretched along the vertical axis for the VP increment, thus approaching the other extreme achiral case of a 3D nanopillar. As a consequence, the CPL discrimination is again reduced in this condition.

Scattering measurements were performed at the single RH particle level for each VP, as a function of circular polarization (Figure 2-c). The circular scattering spectra were collected with a multifunctional confocal setup (see Experimental Section for details). The configuration of the setup allowed to measure all optical signals from a single nanoparticle, when inter-particle distance is larger than 1μm. For the dark field measurements, the single helix nanostructure is illuminated by LCP/RCP light with an oblique incidence angle of 67 degrees. In this case, by placing an objective with numerical aperture around 0.5 above the top of the sample, the reflected beam of the primary light is out of the focus of the objective. Thus, only the pure scattering signal is collected from the top objective. Such configuration is different from a dark-field scheme with condenser applied for the helix nanostructures measurements reported in[21,43–45]. In our case the utilized type of experimental scheme has no limitations related with the substrate transparence and provides the ability to comprehensively study the scattering response even with azimuthal angle dependence.

Very good agreement is obtained with respect to numerical simulations discussed above, confirming that, when the incident light has the same handedness of the helix, the optical modes with low dipole number are more effectively excited. On the contrary, under the excitation of light with opposite handedness, the higher dipole number modes are more efficiently excited, while low frequency dipoles are suppressed. Measurements also quantitatively confirm that the most pronounced CP discrimination occurs for VP of about 550 nm, where scattered intensity of RCP light is nearly doubled as compared to LCP. Furthermore, as a confirmation for fabrication and measurement reliability, spectra in supporting information section in figure S1a, b, c, show that LH enantiomers exhibit exactly opposite chiral scattering responses as compared to RH enantiomers. On the other hand, absence of chiroptical effects is observed in the length/dependent scattering spectra of achiral nanopillars (Figure S1d), grown under the same conditions of helices, with the same helix composition and wire diameter. Therefore, the effective dipolar length and the excitable resonant modes can be finely tuned by varying the helix VP parameter, as experimentally feasible with the growth technique exploited for this work, and this, in turn, controls the coupling with the circularly polarized incident light.

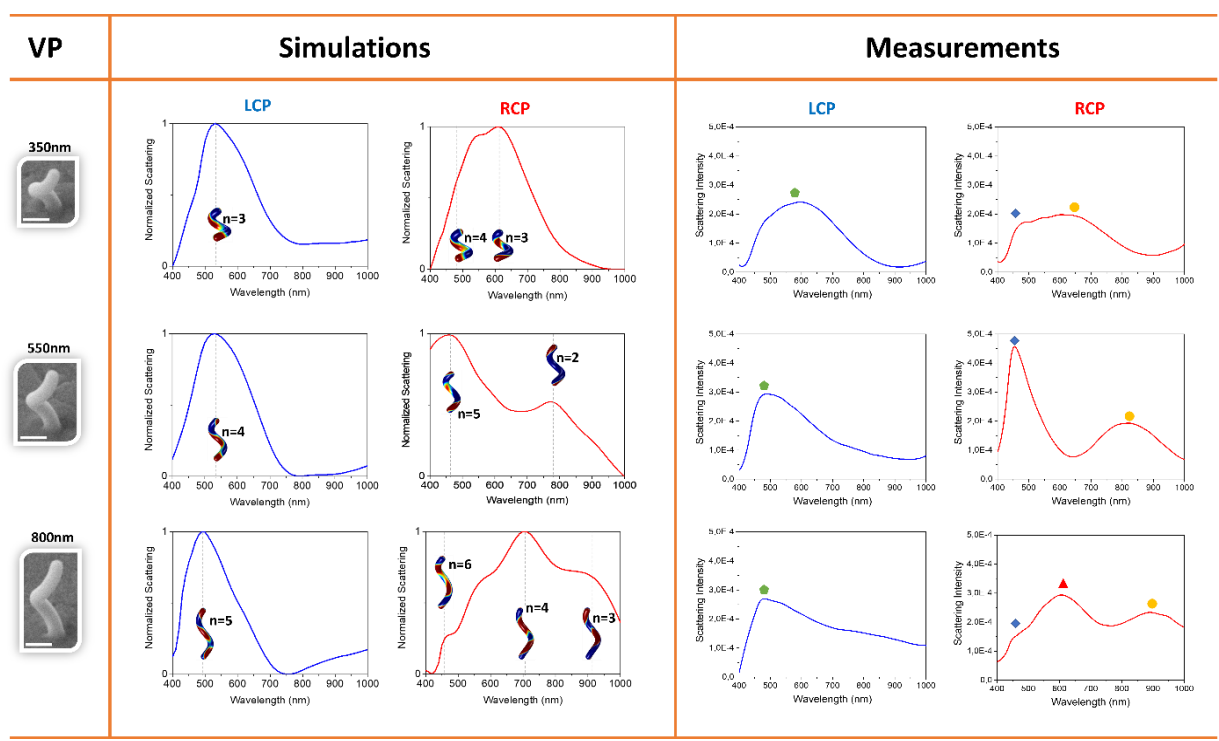

**Figure 2**. Numerical and experimental scattering behavior of single helices as a function of VP (350nm, 550nm and 800nm) and circularly polarized incident light. The normalized scattering spectra are in good agreement for each VP case with the measurements. In particular, the dashed grey lines, in the simulation section, highlight the resonance peaks in correspondence of which we have calculated the electric field distribution on the helix surface to find the associated dipoles number as a function of the incident light polarization. On the other hand, in the measurements side, the colored symbols identify the wavelength position of the main resonances as a function of VP variation and for both the light circular polarizations. In particular, the LCP resonance doesn't suffer the VP variation; differently, for RCP case, while the high n dipole resonance (blue rhomb) remains in the initial position, the low n

dipole peak (yellow circle) shifts towards low frequency region when the VP increases, showing a middle n dipole resonance (red triangle).

Within a periodic array, under normally incident light, coupling between adjacent helices, driven by their mutual distance, is expected to result in a frequency shift of the dipolar resonances, along with enhanced scattering and stronger effective damping [46]. However, it must be taken into account that the out-of-plane helix parameter, investigated above, rules out not only the far field optical resonances of the nanostructures, but also the features of the helix radiation pattern. In particular, numerical calculations reported in Figure 3 show that, while a quasi-axial radiation pattern is exhibited by the single element for all the investigated VPs, for small VP the side lobes can be more tightly confined around the nanostructures (within ~ 500 nm), while they gradually extend up to about 1 micron along the substrate plane as VP increases. Consequently, the possibility to replicate the helix geometry, deterministically and uniformly, throughout a designed array, can allow to further engineer collective interactions via multiple scattering, and realize a CMC, capable to work as a compact and miniaturized polarization control element, potentially integrable within a photonic circuit. It is worth noting that FIB processing, as long as proper calibration steps and pressure stability controls are performed, has demonstrated to be an additive manufacturing solution appropriate to grow large area arrays of complex nano-objects with extremely high reproducibility[47]. Even though further improvements are needed with respect to scalability and time consumption[32], computer aided designs and manufacturing systems ensure reliable multiple element growth in reasonable times. The technique can allocate complex 3D structures as close as few hundreds of nm, being intrinsically limited only by the wire diameter and by the beam probe size. Therefore, collective interactions between neighboring helices can be effectively engineered, even in a closely coupled regime where scattered fields from single element can constructively interfere.

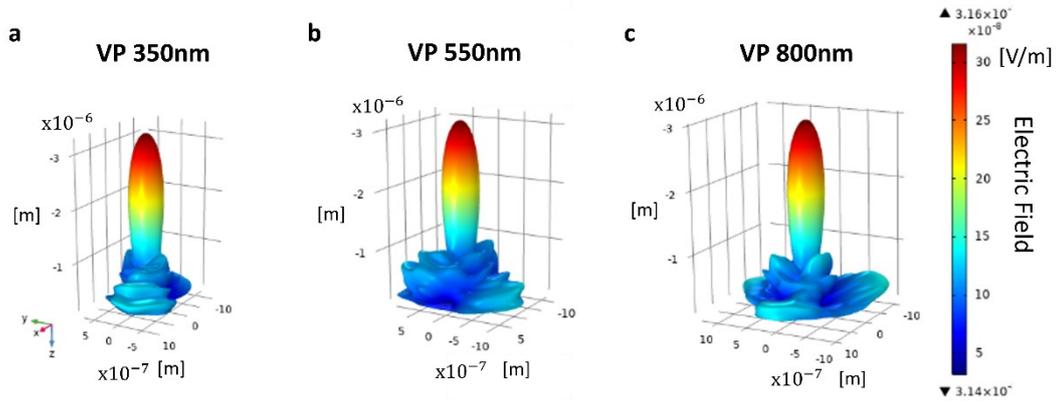

**Figure 3.** Calculated radiation pattern of single nanohelix for RCP incident light at 490 nm with VP of: a) 350 nm, b) 550 nm and c) 800 nm. In all cases a quasi-axial pattern is exhibited, but the side lobe extension progressively increases with VP.

A first insight into the collective behavior can be gained when right-handed nanohelices are periodically arrayed with an in-plane parameter of 700 nm (Figure 4-a), that allows neglectable interaction among the helices, given the side lobes extension of the radiation pattern for a VP of 550 nm (Figure 3-a). The measured transmission spectra from the whole array (see Experimental Section for further details) evidence the presence of optical resonances at the same low frequency and high frequency positions observed from single helix scattering, for all the investigated CP handedness (symbols in Figure 4-b). The measurements also provide an insight into the lossy behavior of our nanostructures, induced by the large k coefficient of the Pt/C composite (Supporting Figure S2), and contributing to the pronounced spectral broadening[48]. Moreover, the effect of the ordered array is the appearance of additional modulations at wavelengths corresponding to diffractive orders of the squared array (in particular (±1;0) and (±1; ±1)), suggesting the possibility to overlap them with the nanostructure resonances towards more complex hybrid states[49,50]. If the lattice is shrunk with LP lower than 700 nm, the mutual scattering among close elements is expected to be enhanced.

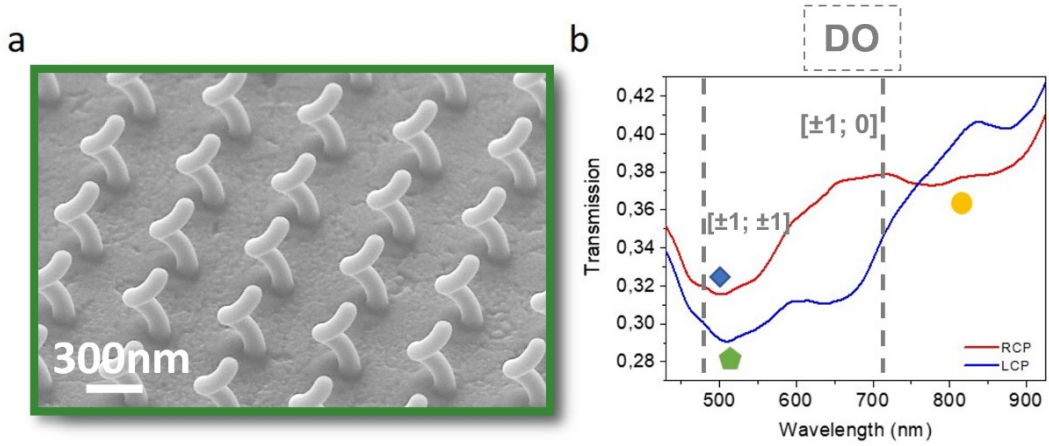

**Figure 4**. a) SEM view of the 10x10 elements CMC with VP=550nm and wide LP of 700nm, in order to reduce mutual helix interaction; b) transmission measurements as a function of CPL for the same array, showing minima (same symbols as in figure 2) corresponding to the dipolar resonances highlighted by scattering experiments and additional modulations which can be inferred to lattice effects (Diffractive orders, DO (±1;0) and (±1; ±1)).

We analyze the results of these effects in the dissymmetry factor (or g-factor, g)[20] of Figure 5, as directly calculated through the normalized difference between the RCP and LCP transmitted light (shown in the supplementary section S3) through the arrays, as follows:

$$g = \frac{2(T_{LCP}-T_{RCP})}{(T_{RCP}+T_{LCP})}.$$ (1)

The g-factor plots show a general redshifting trend with the increase of the crystal constants, more pronounced for the LP effect than for the VP one. In a first approximation, we can assimilate the helix to a multi-stack of split ring resonators (SRRs) and discuss our LP-dependent results in comparison with previous work on arrayed SRRs[51] with geometrical parameters similar to our helices, and excited upon oblique incidence. In that case, for a squared array under linear polarizations, in plane electric dipole coupling and out of plane magnetic dipole coupling were responsible of the redshifts and blueshifts of the transmission resonances when LP is varied, since strong near-field electrostatic and magnetostatic dipole couplings were induced. In our case, by putting the helices in strong proximity (reduced LP value of 430 nm) the total system energy seems to decrease, given the blue-shifted transmission valleys (Figure S3) and related dichroic band extremes and null point (Figure 5-a, b). In particular, the effect of lateral strong interactions among neighbor helices is evident on the short wavelength peak of the bi-signed spectrum. As also observed for arrayed SRRs, it is possible that enhanced scattering and effective super-radiant damping lead to a broadening and weakening of the resonances[52] resulting in the strong reduction of the g-factor intensity for such a short lattice period (LP = 435 nm). On the other hand, a larger spacing between helices induces a spectral redshift which can be ascribed to the magnetic and electric interactions decrement[46], along with a g-factor intensity drop as a consequence of the reduced dipole number per volume (Figure 4-b). The observed results as a function of in-plane lattice parameter anticipate the possibility to excite 3D chiral surface lattice resonances in the VIS range, given that a proper optical impedance matching is realized between substrate and superstrate[53].

Finally, the VP was also finely tuned in the average LP array, within an extended range between 270 nm and 800 nm (Figure 5-c and S3). We note that, for each VP, larger g-factor can be achieved in the blue part of the spectra and that, by increasing the VP value, the $\lambda_g^{Peak}$ (both at short and long wavelength) red shifts (Figure 5-d). The increment of dipolar length accompanied by the increase of scattered field side lobes, modulate the blue-band intensity, which first increases with VP, and after drops down (Figure 5-d), giving an optimal intensity value of about 0.24 for VP in the range between 450 and 500 nm.

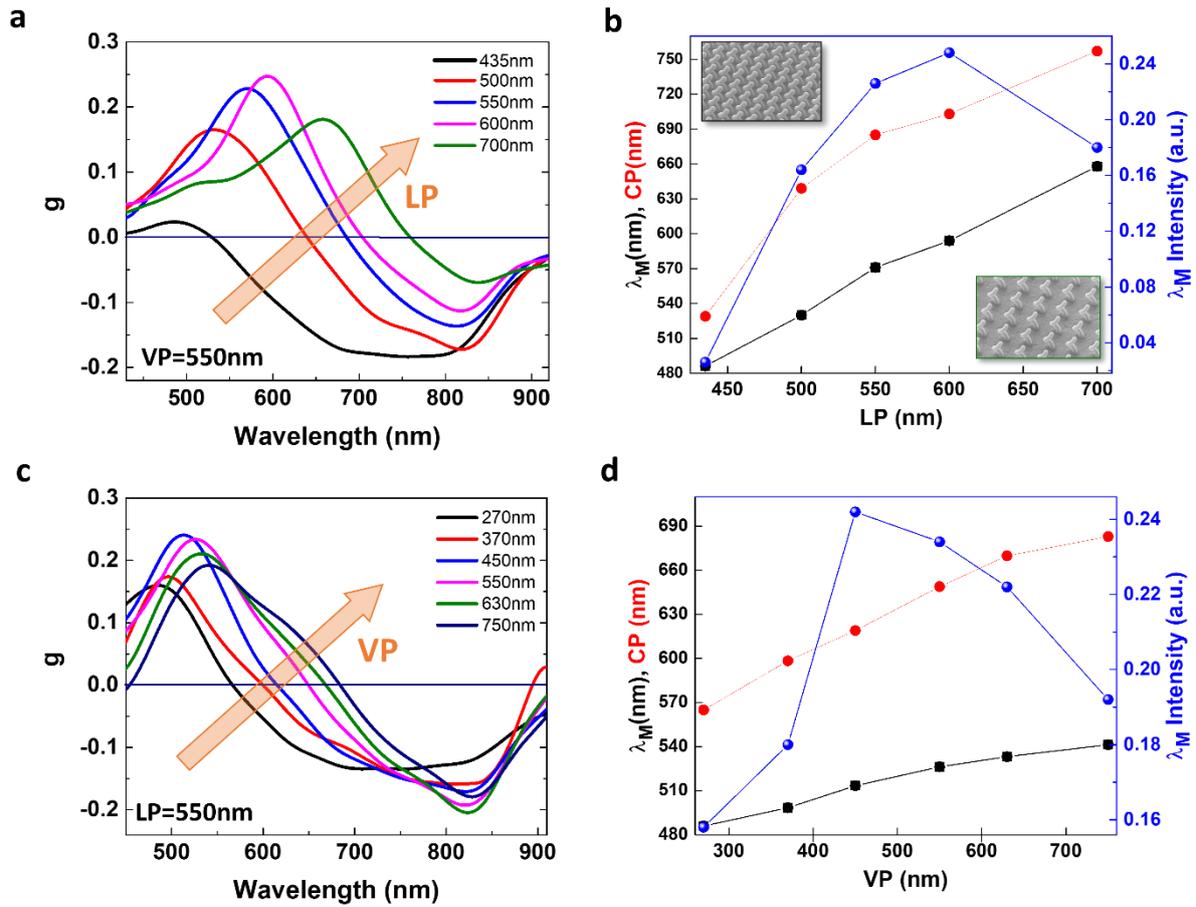

**Figure 5.** a) g-factor spectra retrieved from the measured transmission spectra for the fabricated metacrystals with VP fixed at 550 nm and tuning LP. b) Wavelength peak positions of the maximum peak value $\lambda_M$ as a function of LP (black points) and their relative intensity values (blue line) and their respective null points (red line). c) Dissymmetry factor retrieved from the measured transmission spectra for the fabricated metacrystals with VP fixed at 500 nm and tuning VP. d) Wavelength peak positions of g-factor maximum $\lambda_M$ as a function of VP (black points), their relative intensity values (blue line) and their respective null points (red line).

This analysis demonstrates the existence of a small range of VP values that maximizes the chiroptical properties of the plasmonic helical systems in the visible range, beyond which the structures degenerate toward achiral configurations, which can be the opposite cases of either hollow cylinder or pillar.

To conclude, we have proposed a novel concept of three-dimensional CMC, consisting of nanohelix-shaped metamaterials, with chiroptical properties finely engineered through the in-plane and out-of-plane diffractive coupling. We have shown that under circularly polarized light, different dipolar modes can be excited along the helix arms, depending on the polarization handedness, and on the out-of-plane crystal lattice constant. These modes are the origin of far field optical resonances and of radiation pattern with in-plane extending side lobes. This understanding suggests that collective oscillations between the neighbor helices can be controlled, not only by their mutual distance, but also thanks to a combination of efficient dipole excitation and diffractive coupling matching, allowing to optimize and maximize the overall chiroptical response of the CMC. The tunability of optical response in a broadband spectral range and flexibility of the fabrication process make the CMC a promising platform for creation of miniaturized circular polarizers for photonic circuits[54] which are perspective for a wide set of applications, from optoelectronic devices[55] and displays to integrated quantum chips and biosensors[15].

In particular, given the large chiroptical effects and the metallic nature of the system, novel applications in the emerging field of plasmon-driven chemical reactions[56], involving asymmetric synthesis, as well as in extremely high sensitivity biomolecular detection can be envisioned[14].

**Experimental Section**

*Sample Fabrication.* The samples (both single and nano-helices arrays) have been grown on an ITO-on-glass substrate by means of Focused Ion Beam Induced Deposition technique, by employing a Carl Zeiss Auriga40 Crossbeam FIB/SEM system coupled with a gas injection system (GIS). The Ga+ ion beam is set with very low current, 1pA, and acceleration voltage of 30 keV to perform the helix growth, while trimethyl(methylcyclopentadienyl)platinum(iv) has been used as gaseous precursor. The ion beam breaks the gas molecules locally injected through the nozzle to obtain the controlled growth of the nano helices. During the process, the chamber pressure was kept between $8 \times 10^{-7}$ and $1.06 \times 10^{-6}$ mbar. The step size was set to 10 nm while the dwell time was varied in order to obtain the desired VP. For the nano-helices arrays, the flow conditions were optimized by properly positioning the GIS with respect to the writing zone, together with the application of a dose compensation strategy to control proximity effects[41]. SEM characterization of the single nano-helices and of the arrays was performed by means of a Merlin Zeiss microscope operating in scanning mode.

*Optical characterization.* Transmission spectra were recorded by using an optical microscope Zeiss Axioscope A1 with a home-made confocal system. The sample was back-illuminated with a tungsten lamp, focalized with a condenser with NA<0.1. Then, the transmitted light was collected using a 40x objective lens with NA<0.95. Subsequently, the light was guided through a three lenses system to reconstruct, collimate and refocus the image in the real space. Finally, the light was directed to a CCD camera (Hamamatsu Orca R2) coupled with a 200 mm spectrometer. The real image can be spatially selected with adjustable square slits. The circularly polarized light has been produced using a linear Polarizer (Carl Zeiss, 400-800nm) and a superachromatic waveplate (Carl Zeiss, 400-800nm). The transmission measurements were normalized to the optical response of the substrate. The scattering measurements of the single chiral nanostructure were carried out with a non polarized light from Halogen lamp passing through a Glan linear polarizer and then trough a superachromatic quarter waveplate to generate circularly polarized light. Then the light was focused into the sample by a Mututoyo Plan Apo NIR Objective (M = 10x, NA = 0.26). The oblique incident angle was set at 67 degrees to eliminate the reflection beam from the collecting channel. The scattered light was collected by a Mututoyo Plan Apo NIR Objective (M = 50 x, NA = 0.42) from the top of the sample and then directed to the spectrometer. The distance between single helices was designed to be more than 10μm; in this case the top objective can only capture the scattering signal coming from the selected single plasmonic helix.

*Numerical simulations.* Numerical simulations were performed by exploiting the wave optics module of COMSOL Multiphysics 5.5, by carrying out a frequency domain study of electromagnetic waves propagating among the mere nanostructures, the surrounding medium and perfectly matched layer, on which setting scattering boundary conditions. Here, the scattered wave was considered plane at first order. For far-field calculations the domain is the air surrounding the single helix. The material dispersions used for platinum/carbon alloy are from ref. 6. According to the experimental setup, the structure was excited by a circular polarized plane waves source with azimuthal angle of 60 degrees and placed at enough distance from investigated structure. The scattered signal was collected through an opening of 30 degrees amplitude with respect to the zenith. Wavelength sweeping was performed in a range between 400nm and 1000nm with a step of 10nm, leading to define far field radiation patterns and charge density distribution on the structure surface. The extinction spectra are obtained by comparison among numerical simulations of scattering from structures and dark response ones.

**Acknowledgements**
This work was supported by ''Tecnopolo per la medicina di precisione'' (TecnoMed Puglia) – Regione Puglia: DGR no. 2117 del 21/11/2018 CUP: B84I18000540002. V.T. thanks the Short Term Mobility program of CNR.

**Supporting Information**

## *Three-Dimensional Chiral MetaCrystals*

*Marco Esposito ‖[1*], Mariachiara Manoccio ‖[1], Angelo Leo[1], Massimo Cuscunà[1], Yali Sun[2], Eduard Ageev[2], Dmitry Zuev[2*], Alessio Benedetti[3], Iolena Tarantini[4], Adriana Passaseo[1] and Vittorianna Tasco[1*]*

1. CNR NANOTEC Institute of Nanotechnology, Via Monteroni, 73100 Lecce, Italy
2. ITMO University, Department of Physics and Engineering, 49 Kronverkskiy av., St. Petersburg 197101, Russia
3. D.I.E.T. Department, "Sapienza: Università di Roma", Via Eudossiana 18, I-00184, Rome, Italy
4. Department of Mathematics and Physics Ennio De Giorgi, University of Salento, Via Arnesano, 73100 Lecce.

* Corresponding authors e-mail: *marco.esposito@nanotec.cnr.it; vittorianna.tasco@nanotec.cnr.it, d.zuev@metalab.ifmo.ru*

‖ these authors contributed equally to the work

## S1. Enantiomeric handedness and circularly polarized light-dependence

The interaction between the circularly polarized light and a chiral element leads to a different signal between the two polarization components. In this work, the measured LCP and RCP scattering spectra, are observed for couples of single enantiomers, which have identical sizes (right-handed (RH, red) and left-handed (LH, blue), respectively), fabricated with the same growth condition. When changing the structural handedness, the scattering signal of single right- and left-handed structures shows an opposite chiral behavior, Figures S1 a, b, display the scattering spectra of single 1-loop helices enantiomeric couples (RH and LH, respectively) with 550nm VP height. Their minute mismatches in the optical response of the RH- and LH-helices can be only related to small structural differences fabrication tolerances[1]. For comparison, the scattering spectra for RCP and RCP incident light of an achiral pillar, are plotted in figure S1c. The optical investigation has been performed on a pillar with 550nm height. The spectra show that, the pillar optical response for the two circularly polarized light components, is almost overlapped, indicating the absence of circular polarization-dependence.

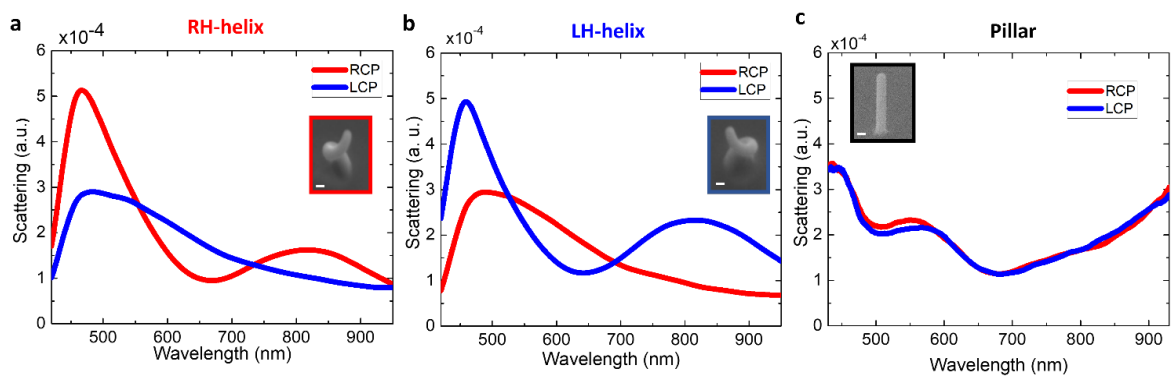

**Figure S1.** LCP and RCP scattering spectra measured for couples of single enantiomers (right-handed and left- handed (blue line), respectively) with VP 550nm, Figure S1a, b, respectively. These spectra show that, for two enantiomeric couples, the chiral behavior is inverted for all the investigated enantiomeric couples. c) Scattering spectra for RCP and RCP incident light on a pillar with 550 nm height, framed in black in the figure. The spectra show that, for a non chiral element, like the pillar, the optical response for the two circularly polarized light components, is overlapped. The insets represent the enantiomeric helix couples and the achiral nanopillar with scale bar 100nm.

## S2. Compositional profile of Platinum-based helix systems

Platinum based-helical nanostructures grown by Focused ion beam induced deposition, exhibit a compositional profile made of a platinum-carbon alloy with platinum nanograins (with averaged size of 5nm), embedded in an amorphous carbon matrix with a volume percentage of 50% Pt and 45% C and a low residual Ga content, <5%. Figure S2a is a schematical representation of an helix section where the carbon matrix is the green disk, while the platinum grains are the blue circles. The optical dispersion has been numerically calculated using a finite difference time-domain based software (Lumerical FDTD Solutions), considering the material composition retrieved by Transmission Electron Microscope and Energy Dispersive X-ray Spectroscopy in[2]. Owed to the complex compositional profile of the Pt helix, an effective medium approach can be applied. Particularly, FDTD analysis have been carried out for a C host matrix, enclosing randomly distributed spheroids (representing Ga clusters and platinum nanoparticles) in a large FDTD box with periodic boundary conditions and illuminating the system with a plane wave with normal incidence. The reflection coefficients (r) of the overall compound are retrieved and the effective refractive index is calculated as:

$$n(\lambda) = n_{host}(\lambda) \cdot \frac{1 - r(\lambda)}{1 - r(\lambda)}$$

where $n_{host}$ is air refractive index and r is the complex reflection coefficient. Using the dispersion values of Pt from[3], Ga from[4] and Carbon from[5], the dispersion curves are retrieved and displayed in Figure S2b.

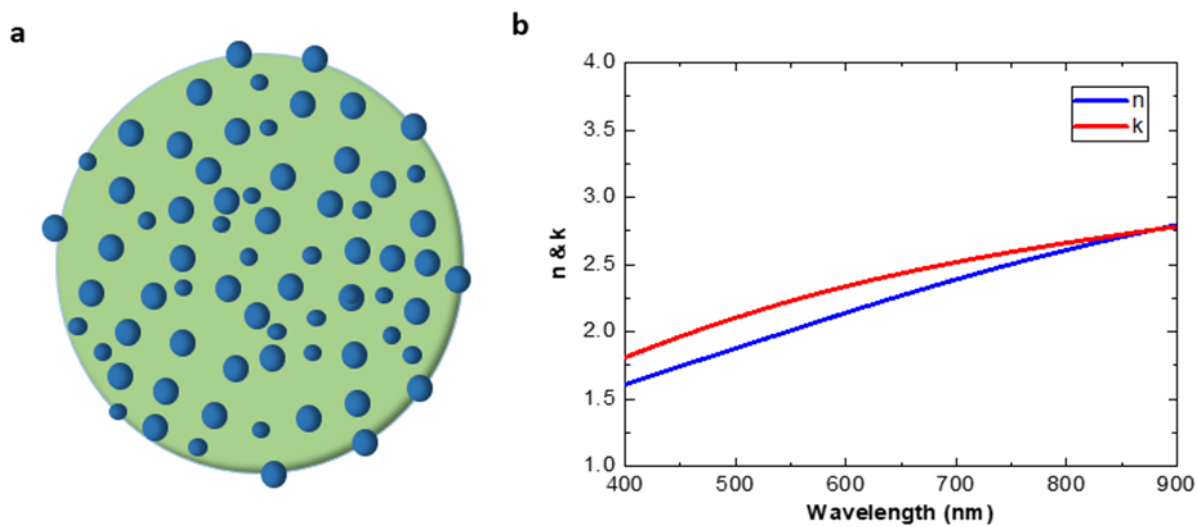

**Figure S2.** a) Schematic view of the wire cross section which shows that the structure is composed by Pt nanograins embedded in an amorphous C matrix. The blue points correspond to Pt grains, while the amorphous carbon matrix is represented by the green disk. d) Analytical dispersion values (n-blue line, k-red line) calculated for the Pt-based helix.

**S3. Chiral Metacrystals as a function Array of Pt helices as a function of the unit cell**

The transmission spectra measured for chiral metacrystals, as well as the SEM images depicting the metacrystal morphology, by tuning the unit cell, are observed in figure S3.

The overall trend shows a redshift of all the spectral features, when LP is fixed and varying VP, as observed for their circular dichroism spectra. When tuning the lattice periodicity, instead , beside the redshift of the main spectral range, the measured transmission spectra also exhibit an enlargement of the circular birefringence, together with the sharpening of other peaks corresponding to the diffractive orders owed to the in-plane interactions among the helices.

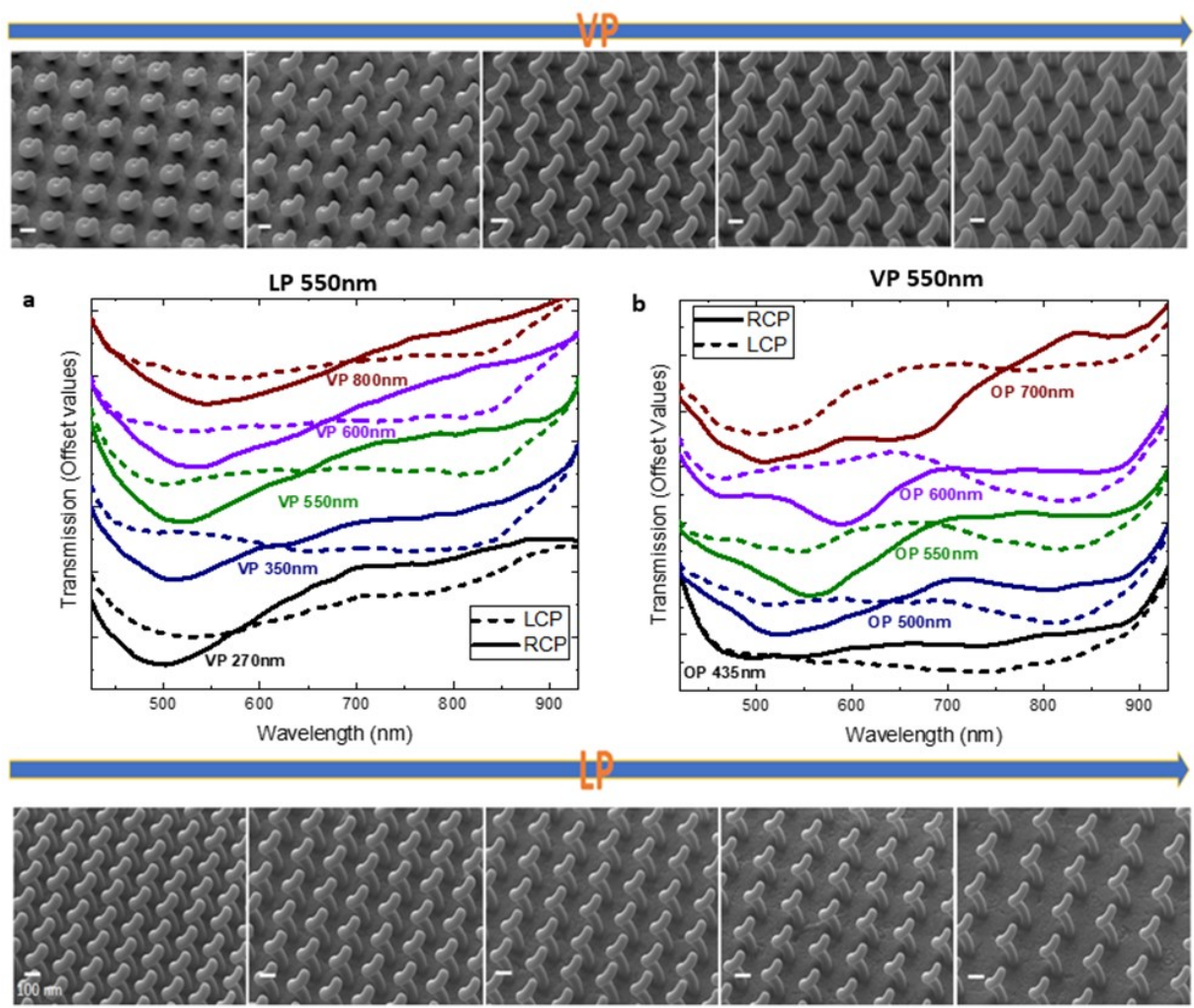

**Figure S1.** SEM Images of the built chiral metacrystals composed by right-handed helices with different periodicities: Top row: LP is fixed at 500nm while VP varies VP of 270nm, 350nm, 550nm, 600nm, 800nm , from left to right, respectively. Bottom Row: VP is fixed at 550nm with variable LPs of 435nm, 500nm, 550nm, 600nm, 700nm, from left to right, respectively. The scale bar is 100nm. In the central panel there is a schematic representation of the geometrical parameter of the unit cell: The Vertical

Period (VP), External Diameter (ED), the wire Diameter (WD), the Lateral Period (LP). At the center: Transmission spectra for the Metacrystals imaged above a) With LP fixed at 550nm and varying VP and b) With VP fixed at 550nm and varying LP.